\documentclass[12pt]{article}%
\usepackage{amsmath,latexsym}
\usepackage{graphicx}
\usepackage{amsmath}
\usepackage{amsfonts}
\usepackage{amssymb}%
\begin{document}

\title{{Conformal symmetry wormholes and the null energy
      condition}}
   \author{
Peter K. F. Kuhfittig*\\
\footnote{E-mail: kuhfitti@msoe.edu}
 \small Department of Mathematics, Milwaukee School of
Engineering,\\
\small Milwaukee, Wisconsin 53202-3109, USA}

\date{}
 \maketitle

\begin{abstract}\noindent
In this paper we seek a relationship between
the assumption of conformal symmetry and the
exotic matter needed to hold a wormhole open.
By starting with a Morris-Thorne wormhole
having a constant energy density, it is shown
that the conformal factor provides the extra
degree of freedom sufficient to account for
the exotic matter.  The same holds for
Morris-Thorne wormholes in a
noncommutative-geometry setting.  Applied
to thin shells, there would exist a radius
that results in a wormhole with positive
surface density and negative surface pressure
and which violates the null energy condition
on the thin shell. \\

\end{abstract}

\section{Introduction}\label{S:Introduction}

Wormholes are handles or tunnels in spacetime
that connect different regions of our Universe
or completely different universes.  That
wormholes could be macroscopic structures
allowing interstellar travel was first
proposed by Morris and Thorne \cite{MT88}.
With the Schwarzschild line element in mind,
such a wormhole could be described by the
static and spherically symmetric line element:
\begin{equation}\label{E:line1}
ds^{2}=-e^{2\Phi(r)}dt^{2}+\frac{dr^2}{1-b(r)/r}
+r^{2}(d\theta^{2}+\text{sin}^{2}\theta\,
d\phi^{2}),
\end{equation}
using units in which $c=G=1$.  Here
$\Phi=\Phi(r)$ is referred to as the
\emph{redshift function}, which must be
everywhere finite to avoid an event horizon.
The function $b=b(r)$ is called the
\emph{shape function} since it helps to
determine the spatial shape of the
wormhole \cite{MT88}.  The spherical 
surface $r=r_0$ is the \emph{throat} 
of the wormhole.  At the throat, $b=b(r)$
must satisfy the following conditions:
$b(r_0)=r_0$, $b(r)<r$ for $r>r_0$, and
$b'(r_0)\le 1$, usually called the
\emph{flare-out condition}.  This condition
can only be satisfied by violating the null
energy condition, defined as follows: for
the stress-energy tensor $T_{\alpha\beta}$,
we must have
\begin{equation}\label{E:nullvectors}
  T_{\alpha\beta}\mu^{\alpha}\mu^{\beta}
  \ge 0
\end{equation}
for all null vectors.  By Ref. \cite{MT88},
the violation is equivalent to the condition
\begin{equation}
  \frac{b'(r_0)-b(r_0)/r_0}
     {2[b(r_0)]^2}<0.
\end{equation}
For a Morris-Thorne wormhole, matter that
violates the null energy condition is called
``exotic."

In this paper, we are going to seek a
relationship between exotic matter and
conformal symmetry, by which is meant the
existence of a conformal Killing vector
$\xi$ defined by the action of
$\mathcal{L_{\xi}}$ on the metric tensor
\begin{equation}
  \mathcal{L_{\xi}}g_{\mu\nu}=\psi(r)\,g_{\mu\nu};
\end{equation}
here $\mathcal{L_{\xi}}$ is the Lie derivative
operator and $\psi(r)$ is the conformal factor.

It is shown that $\psi(r)$ provides the extra
degree of freedom to account for the exotic
matter for certain types of wormholes.  Applied
to thin-shell wormholes, an appropriate choice
of the radius avoids the usual negative surface
density typical of thin shells.

\section{Conformal Killing vectors}
    \label{S:Killing}
We assume in this paper that our static spherically
symmetric spacetime admits a one-parameter group of
conformal motions, i.e., motions along which
the metric tensor of a spacetime remains invariant
up to a scale factor.  Equivalently, there exist
conformal  Killing vectors such that
\begin{equation}\label{E:Lie}
   \mathcal{L_{\xi}}g_{\mu\nu}=g_{\eta\nu}\,\xi^{\eta}
   _{\phantom{A};\mu}+g_{\mu\eta}\,\xi^{\eta}_{\phantom{A};
   \nu}=\psi(r)\,g_{\mu\nu},
\end{equation}
where the left-hand side is the Lie derivative of the
metric tensor and $\psi(r)$ is the conformal factor
\cite{MM96, BHL07}.  The vector $\xi$ generates the
conformal symmetry and the metric tensor $g_{\mu\nu}$
is conformally mapped into itself along $\xi$.  This
type of symmetry has proved to be effective in
describing relativistic stellar-type objects
\cite{HPa, HPb}.  Moreover, conformal symmetry has
led to new solutions, as well as to new geometric
and kinematical insights \cite{MS93, Ray08, fR10, fR12b}.
Two earlier studies assumed \emph{non-static}
conformal symmetry \cite{BHL07, BHL08}.

To study the effect of conformal symmetry, it is
convenient to use an alternate form of the metric
\cite{RRKKI}:
\begin{equation}\label{E:line2}
   ds^2=- e^{\nu(r)} dt^2+e^{\lambda(r)} dr^2
   +r^2( d\theta^2+\text{sin}^2\theta\, d\phi^2).
\end{equation}
Using this line element, the Einstein field
equations become

\begin{equation}\label{E:Einstein1}
e^{-\lambda}
\left(\frac{\lambda^\prime}{r} - \frac{1}{r^2}
\right)+\frac{1}{r^2}= 8\pi \rho,
\end{equation}

\begin{equation}\label{E:Einstein2}
e^{-\lambda}
\left(\frac{1}{r^2}+\frac{\nu^\prime}{r}\right)-\frac{1}{r^2}=
8\pi p_r,
\end{equation}

\noindent and

\begin{equation}\label{E:Einstein3}
\frac{1}{2} e^{-\lambda} \left[\frac{1}{2}(\nu^\prime)^2+
\nu^{\prime\prime} -\frac{1}{2}\lambda^\prime\nu^\prime +
\frac{1}{r}({\nu^\prime- \lambda^\prime})\right] =8\pi p_t.
\end{equation}
Here $\rho$ is the energy density, while $p_r$ and $p_t$
are the radial and transverse pressures, respectively.
Eq. (\ref{E:Einstein3}) could actually be obtained from
the conservation of the stress-energy tensor, i.e.,
$T^{\mu\nu}_{\phantom{\mu\nu};\nu}=0$.  So we need to use
only Eqs. (\ref{E:Einstein1}) and (\ref{E:Einstein2}).

The subsequent analysis can be simplified somewhat
by following Herrera and Ponce de Le\'{o}n \cite{HPa}
and restricting the vector field by requiring
that $\xi^{\alpha}U_{\alpha}=0$, where $U_{\alpha}$
is the four-velocity of the perfect fluid
distribution.  As a result, fluid flow lines are
mapped conformally onto fluid flow lines.  The
assumption of spherical symmetry then implies that
$\xi^0=\xi^2=\xi^3=0$ \cite{HPa}.  Eq. (\ref
{E:Lie}) now yields the following results:
\begin{equation}\label{E:sol1}
    \xi^1 \nu^\prime =\psi,
\end{equation}
\begin{equation}\label{E:sol2}
   \xi^1  = \frac{\psi r}{2},
\end{equation}
and
\begin{equation}\label{E:sol3}
  \xi^1 \lambda ^\prime+2\,\xi^1 _{\phantom{1},1}=\psi.
\end{equation}
From Eqs. (\ref{E:sol1}) and (\ref{E:sol2}), we then
obtain
\begin{equation} \label{E:gtt}
   e^\nu  =c_1 r^2,
\end{equation}
which, combined with Eq. (\ref{E:sol3}), produces
another important result:
\begin{equation}\label{E:grr}
   e^\lambda  = \left(\frac {c_2} {\psi}\right)^2;
\end{equation}
$c_1$ and $c_2$ are integration constants.

The field equations (\ref{E:Einstein1}) and
(\ref{E:Einstein2}) can be rewritten as follows:
\begin{equation}\label{E:E1}
\frac{1}{r^2}\left(1 - \frac{\psi^2}{c_2^2}
\right)-\frac{(\psi^2)^\prime}{c_2^2r}= 8\pi \rho
\end{equation}
and
\begin{equation}\label{E:E2}
\frac{1}{r^2}\left( \frac{3\psi^2}{c_2^2}-1
\right)= 8\pi p_r.
\end{equation}
It now becomes apparent that $c_2$ is merely a
scale factor in Eqs. (\ref{E:grr})-(\ref{E:E2});
so we may assume that $c_2=1$.  The constant
$c_1$, however, has to be obtained from the
junction conditions, the need for which can be
seen from Eq. (\ref{E:gtt}): since our
wormhole spacetime is not asymptotically flat,
the wormhole material must be cut off at some
$r=a$ and joined to an exterior Schwarzschild
solution,
 \begin{equation}
ds^{2}=-\left(1-\frac{2M}{r}\right)dt^{2}
+\frac{dr^2}{1-2M/r}
+r^{2}(d\theta^{2}+\text{sin}^{2}\theta\,
d\phi^{2}).
\end{equation}
so that $e^{\nu(a)}=c_1a^2=1-2M/a$, whence
\begin{equation}\label{E:cutoff}
   c_1=\frac{1-2M/a}{a^2},
\end{equation}
where $M$ is the mass of the wormhole as seen
by a distant observer.  It also follows that
$b(a)=2M$.

\section{Wormhole structure}

To simplify the analysis in the next section,
we will assume that the energy density $\rho$
is constant and that the wormhole material
is confined to the spherical shell $r_0\le
r\le a$, where $a$ is the cut-off in Eq.
(\ref{E:cutoff}).  (This form of $\rho(r)$
was also assumed by Sushkov \cite{sS05} in
his discussion of wormholes supported by
phantom energy.)  So from Eq. (\ref{E:E1})
(with $c_2=1$), we obtain the following
differential equation:
\begin{equation}\label{E:diff1}
   \frac{1}{r^2}(1-\psi^2)-\frac{(\psi^2)'}{r}=
   8\pi\rho_0,\quad r_0\le r\le a.
\end{equation}
After multiplying by $r$ and rearranging,
we obtain
\begin{equation}\label{E:diff2}
   (\psi^2)'+\frac{1}{r}\psi^2=\frac{1}{r}
   -8\pi\rho_0r.
\end{equation}
This equation is linear in $\psi^2$ and can
readily be solved to obtain
\begin{equation}\label{E:solution}
   \psi^2=1-\frac{8}{3}\pi\rho_0 r^2+
   \frac{c}{r},
\end{equation}
where $c$ is a constant of integration.
From Eqs. (\ref{E:line1}) and (\ref{E:line2}),
we get $e^{-\lambda}=1-b(r)/r$, whence
\begin{equation}\label{E:shape1}
   b(r)=r(1-\psi^2).
\end{equation}
The requirement $b(r_0)=r_0$ now implies that
$\psi^2(r_0)=0$.  So by Eq. (\ref{E:solution}),
\begin{equation}\label{E:c}
  c=\frac{8}{3}\pi\rho_0r_0^3-r_0,
\end{equation}
and from Eq. (\ref{E:shape1}),
\begin{equation}\label{E:shape2}
   b(r)=r\left(\frac{8}{3}\pi\rho_0r^2
   -\frac{8}{3}\pi\rho_0r_0^3\frac{1}{r}
   +\frac{r_0}{r}\right), \quad r_0\le r\le a.
\end{equation}

Next, to meet the flare-out condition, we
require that
\begin{equation}\label{E:bprime}
   \left. b'(r)=\frac{8}{3}\pi\rho_0
   (3r^2)\right |_{r=r_0}=\frac{8}{3}
   \pi\rho_0 (3r_0^2)<1,
\end{equation}
which implies that
\begin{equation}\label{E:rhozero}
   \rho_0<\frac{1}{8\pi r_0^2}.
\end{equation}

\section{The null energy condition}

Returning to the null energy condition
(\ref{E:nullvectors}),
$T_{\alpha\beta}\mu^{\alpha}\mu^{\beta}\ge 0$,
observe that for the radial outgoing null
vector $(1,1,0,0)$, we obtain $\rho+p_r<0$
whenever the condition is violated.  As noted
earlier, for a Morris-Thorne wormhole, matter
that violates the null energy condition is
usually called ``exotic."  Moreover, we saw
in the previous section that $b'(r_0)<1$
whenever $\rho_0<1/(8\pi r_0^2)$.  The extra
assumption of conformal symmetry now yields
by Eq. (\ref{E:E2})
\begin{equation}\label{E:pr}
   p_r=\frac{1}{8\pi}\frac{1}{r^2}
   [3\psi^2(r)-1]
\end{equation}
and by Eq. (\ref{E:rhozero}), as expected,
\begin{equation}\label{E"null1}
   \rho +p_r|_{r=r_0}=\rho_0+\frac{1}{8\pi}
   \frac{1}{r_0^2}[3\psi^2(r_0)-1]<0
\end{equation}
since $\psi^2(r_0)=0$.  This result suggests
that the assumption of conformal symmetry
helps explain the violation of the null
energy condition by accounting for the
exotic matter.  More precisely, while the
physical requirements impose some severe
constraints on the geometry, they do not
determine the conformal factor $\psi(r)$.
Such a determination depends on other
important geometric notions such as shape
characteristics and shape deformations
that arise in various fields such as
computer graphics.  (For further
discussion, see Refs. \cite{jB76, HM00,
mB08, WG10, AKZ}.)  These geometric
factors provide an extra degree of
freedom via $\psi(r)$ that is not
available for the usual Morris-Thorne
wormholes.

The case for eliminating (in the above
sense) exotic matter in certain cases
can be strengthened in the context of
noncommutative geometry, which replaces
point-like structures by smeared objects.
The smearing effect can be accomplished
by assuming that the energy density of
a static and spherically symmetric and
particle-like gravitational source has
the form \cite{LL12, NM08,
KG14, pK15}
\begin{equation}\label{E:rho1}
  \rho(r)=\frac{M\sqrt{\theta}}
     {\pi^2(r^2+\theta)^2}.
\end{equation}
Here the mass $M$ is diffused throughout
the region of linear dimension
$\sqrt{\theta}$ due to the uncertainty.
Observe that $\rho +p_r$ now becomes
\begin{equation}\label{E:null2}
   \rho(r)+p_r(r)=\frac{M\sqrt{\theta}}
   {\pi^2(r^2+\theta)^2}+\frac{1}{8\pi}
   \frac{1}{r^2}[3\psi^2(r)-1]
\end{equation}
and
\begin{equation}\label{E:null3}
   \rho(r_0)+p_r(r_0)=\frac{M\sqrt{\theta}}
   {\pi^2(r_0^2+\theta)^2}-\frac{1}{8\pi}
   \frac{1}{r_0^2}<0
\end{equation}
since $\sqrt{\theta}\ll 1$.  So the violation
of the null energy condition can be attributed
to a combination of physical and geometric
factors.

\section{Thin-shell wormholes}

Using the now standard cut-and-paste technique,
a thin-shell wormhole is constructed by taking
two copies of a Schwarzschild spacetime and
removing from each the four-dimensional region
\begin{equation}
   \Omega= \{r\le a\,|\,a>2M\},
\end{equation}
where $a$ is a constant \cite{PV95}.  By
identifying the boundaries, i.e., letting
\begin{equation}
   \partial\Omega= \{r=a\,|\,a>2M\}
\end{equation}
we obtain a manifold that is geodesically
complete, while possessing two asymptotically
flat regions connected by a wormhole.  The
throat is the surface $\partial\Omega$.

Since the shell is infinitely thin, the radial
pressure is zero.  So if the surface density
is denoted by $\sigma$, then $\sigma +p_r<0$
implies that $\sigma$ is negative.  The goal
in this section is to show that under the
assumption of conformal symmetry, $\sigma$
can be positive.  The null energy condition
will then be violated on the thin shell itself,
even though it is met for the radial outgoing
null vector $(1,1,0,0)$.

To that end, let us consider the surface
stresses.  So we need to recall the
Lanczos equations
\cite{fL04}
\begin{equation}\label{E:sigma1}
     \sigma=-\frac{1}{4\pi}\kappa^{\theta}_
     {\phantom{\theta}\theta}
\end{equation}
and
\begin{equation}
   \mathcal{P}=\frac{1}{8\pi}(\kappa^{\tau}
   _{\phantom{\tau}\tau}+\kappa^{\theta}
   _{\phantom{\theta}\theta}),
\end{equation}
where $\kappa_{ij}=K^{+}_{ij}-K^{-}_{ij}$ and
$K_{ij}$ is the extrinsic curvature.  According to
Ref. \cite{fL04},
\begin{equation*}
   \kappa^{\theta}_{\phantom{\theta}\theta}=\frac{1}{a}
   \sqrt{1-\frac{2M}{a}}-\frac{1}{a}
      \sqrt{1-\frac{b(a)}{a}}.
\end{equation*}
So by Eq. (\ref{E:sigma1}),
\begin{equation}\label{sigma2}
   \sigma=-\frac{1}{4\pi a}
   \left(\sqrt{1-\frac{2M}{a}}-
      \sqrt{1-\frac{b(a)}{a}}\right).
\end{equation}
In view of the assumption $b(a)=2M$, one could
reasonably expect that $\sigma=0$.  However, part
of the junction formalism is to assume that the
junction surface $r=a$ is an infinitely thin
surface having a nonzero density that may be
positive or negative. So we have instead, $b(a)
\approx 2M$.  Again following Ref \cite{fL04},
\[
   K^{\tau\,+}_{\phantom{\tau}\tau}=\frac{M/a^2}
   {\sqrt{1-2M/a}}
\]
and
\[
   K^{\tau\,-}_{\phantom{\tau}\tau}=
   \Phi'(a)\sqrt{1-\frac{b(a)}{a}}.
\]
The surface pressure is therefore given by
\begin{equation}\label{E:pressure1}
   \mathcal{P}=\frac{1}{8\pi}\left[\frac{M/a^2}
   {\sqrt{1-2M/a}}-\Phi'(a)
   \sqrt{1-\frac{b(a)}{a}}
   +\frac{1}{a}\sqrt{1-\frac{2M}{a}}
   -\frac{1}{a}\sqrt{1-\frac{b(a)}{a}}\right].
\end{equation}

To apply these ideas, let us describe a thin
shell by starting with a typical shape
function and letting $r_0$ be arbitrarily close
to $a$, again taken to be the cut-off.  This
cut-off results in the redshift function $\nu=
2\Phi=\text{ln}\,c_1r^2$ by Eq. (\ref{E:gtt});
here $c_1=(1-2M/a)/a^2$ from Eq. (\ref{E:cutoff}).
As a result, $\Phi'(a)=1/a$.

Since $r_0$ is arbitrarily close to $a$, the
junction surface itself becomes the thin shell
with $r=a$.  So if $b(r)$ is a typical shape
function, then, ignoring the cut-off for now,
we have $b(r_0)=r_0$, $b(r)<r$ for
$r>r_0$, and $\text{lim}_{r\rightarrow\infty}
b(r)/r=0$.  So $b(a)/a$ assumes all values
on the interval $(0,1]$ at least once.
It follows that $a$ can be chosen to yield any
desired value for $b(a)/a$ in the interval
$(0,1]$.

Returning now to Eq. (\ref{E:pressure1}), since
$\Phi'(a)=1/a$,
\begin{equation}\label{E:P}
   \mathcal{P}=\frac{1}{8\pi}\left(
   \frac{\frac{M}{a^2}-\frac{1}{a}
   \sqrt{1-\frac{b(a)}{a}}\sqrt{1-\frac{2M}{a}}}
   {\sqrt{1-\frac{2M}{a}}}\right)+\frac{1}{2}
   \frac{1}{4\pi a}\left(\sqrt{1-\frac{2M}{a}}-
   \sqrt{1-\frac{b(a)}{a}}\right).
\end{equation}
Observe that the last term is equal to
$\frac{1}{2}\sigma$ in absolute value.  Moreover,
if $\sigma$ is positive, then the last term is
negative.

To see how $\mathcal{P}$ can be negative (while
$\sigma >0$), let us assume for a moment that
$b(a)=2M$, making $\sigma =0$.  Then
\begin{equation}\label{E:sigma0}
   \frac{M}{a^2}-\frac{1}{a}\sqrt{1-\frac{b(a)}{a}}
   \sqrt{1-\frac{2M}{a}}
   =\frac{\frac{1}{2}b(a)}{a^2}-\frac{1}{a}
        \left(1-\frac{b(a)}{a}\right)
   =\frac{1}{a}\left(-1
   +\frac{3}{2}\frac{b(a)}{a}\right)=0
\end{equation}
whenever $b(a)/a=2/3$, but if $b(a)\approx 2M$,
then we only have $b(a)/a\approx 2/3$.  As already
noted, since $b(a)/a$ assumes all values on the
interval  $(0,1]$ at least once, such a choice
can be made.  Referring to Eq. (\ref{E:sigma0}),
it now follows that
\[
   \frac{1}{a}\left(-1+\frac{3}{2}
      \frac{b(a)}{a}\right)<0
\]
whenever $b(a)/a\lesssim 2/3$.  More precisely,
suppose $b(a)/a=2/3-k_a$ for some $k_a>0$.
Then with Eq. (\ref{E:P}) in mind, consider
a value for $k_a$ for which
\begin{equation}\label{E:ineq}
   \frac{1}{8\pi}\frac{1}{\sqrt{1-2M/a}}
   \frac{1}{a}
   \left[-1+\frac{3}{2}\left(\frac{2}{3}-k_a
   \right)\right]<\frac{1}{8\pi}\frac{2}{a}
   \left(\sqrt{1-\frac{2M}{a}}-
      \sqrt{1-\frac{b(a)}{a}}\right);
\end{equation}
observe that the right side is equal to
$\sigma$ in absolute value.  Solving for
$k_a$ shows that such a $k_a$ exists:
\begin{equation}
   k_a>-\frac{4}{3}
   \left(\sqrt{1-\frac{2M}{a}}-
   \sqrt{1-\frac{b(a)}{a}}\right)
      \sqrt{1-\frac{2M}{a}}>0.
\end{equation}
As a result, the first term on the 
right-hand side of Eq. (\ref{E:P}) is 
negative and bounded away from zero, 
while the second term can be arbitrarily 
close to zero.  The same is true for  
\begin{equation}
   \mathcal{P}+ \sigma=\frac{1}{8\pi}\left(
   \frac{\frac{M}{a^2}-\frac{1}{a}
   \sqrt{1-\frac{b(a)}{a}}\sqrt{1-\frac{2M}{a}}}
   {\sqrt{1-\frac{2M}{a}}}\right)-\frac{1}{2}
   \frac{1}{4\pi a}\left(\sqrt{1-\frac{2M}{a}}-
   \sqrt{1-\frac{b(a)}{a}}\right).
\end{equation}
Hence
\begin{equation}\label{E:nullshell}
   \sigma +\mathcal{P}<0,
\end{equation}
which was to be shown.

\section{Conclusion}

The purpose of this paper is to seek a relationship
between the assumption of conformal symmetry and
the exotic matter needed to hold a wormhole open.
It was concluded that the conformal factor $\psi(r)$
provides an extra degree of freedom sufficient to
account for the exotic matter for certain types
of wormholes, those having a constant energy
density on the spherical shell $r_0\le
r\le a$ and wormholes in a noncommutative-geometry
setting.  The extra degree of freedom does not
exist for the usual Morris-Thorne wormholes.

Applied to thin shells, the assumption of
conformal symmetry implies that the surface
density can be positive and the surface pressure
negative for some radius $r=a$ and that the null
energy condition is violated on the thin shell.

\end{document}